\documentclass{aa}
\usepackage{graphicx}

\begin{document}

\title{Gamma-ray emission from Wolf-Rayet binaries}

\author{P. Benaglia\inst{1,}\thanks{Member of  Carrera del Investigador,
CONICET},
  G.~E. Romero\inst{1,}$^{\star}$ }

\offprints{P. Benaglia: paula@lilen.fcaglp.unlp.edu.ar }

\institute{$^1$ Instituto Argentino de Radioastronom\'{\i}a,
   C.C.5, (1894) Villa Elisa, Buenos Aires, Argentina}

\date{\today}

\abstract{In the colliding wind region of early-type binaries,
electrons can be accelerated up to relativistic energies
displaying power-law spectra, as demonstrated by the detection of
non-thermal radio emission from several WR+OB systems. The
particle acceleration region, located between the stars, is
exposed to strong photon fields in such a way that inverse Compton
cooling of the electrons could result in a substantial high-energy
non-thermal flux. In particular cases, the ratio of the energy
densities of magnetic to photon fields in the colliding wind
region will determine whether a given source can produce or not
significant gamma-ray emission. We present here a study of the
binaries WR 140, WR 146, and WR 147 in the light of recent radio
and gamma-ray observations. We show that with reasonable
assumptions for the magnetic field strength WR 140 can produce the
gamma-ray flux from the EGRET source 3EG J2022+4317. WR 146 and WR
147 are below the detection threshold, but forthcoming instruments
like INTEGRAL and GLAST could detect non-thermal emission from
them.
 }

\titlerunning{Gamma-ray emission from WRs}
\authorrunning{Benaglia \& Romero}
\maketitle

\markboth{P. Benaglia and G.E. Romero: Gamma-ray emission from
WRs}{}

\section{Introduction} \label{introduction}

Early-type stars present strong supersonic winds that are
responsible for significant mass loss rates, which in Wolf-Rayet
(WR) stars can reach values close to $10^{-4}$ $M_{\sun}$
yr$^{-1}$ (Abbott et al. \cite{abbott86}, Leitherer et al.
\cite{leitherer97}). These winds interact with the interstellar
medium sweeping up the ambient material, creating cavities or
bubbles (e.g. Benaglia \& Cappa \cite{benaglia99}), and forming
strong shock fronts.

At the terminal shocks, with typical velocities of thousands of km
s$^{-1}$, locally-injected supra-thermal particles could be
accelerated up to relativistic energies and power-law
distributions (e.g. Cass\'e \& Paul \cite{casse80}, V\"olk \&
Forman \cite{volk82}). Energetic particles from the star lose too
much energy in the expanding wind before reaching the shock to be
efficiently accelerated, but it has been suggested that partial
re-acceleration during their travel, produced by multiple shocks
from line-driven instabilities in the inner wind region, could
compensate the adiabatic losses (White \cite{white85}). In any
case, if a continuous injection of supra-thermal protons or ions
can be sustained, the subsequent interaction of the relativistic
hadrons with ambient atoms will produce gamma-ray emission through
the neutral pion chain: $p+p\rightarrow\pi^{0}+ X$,
$\pi^{0}\rightarrow \gamma+\gamma$.

For typical densities in the interstellar medium, however, the
expected gamma-ray luminosity from isolated massive stars at
energies $E>100$ MeV is in the range $10^{32-33}$ erg s$^{-1}$
(Benaglia et al. \cite{benaglia01}), too low to be detected at
standard distances, by instruments like the Energetic Gamma-Ray
Experiment Telescope (EGRET) of the recently terminated Compton
Gamma-Ray Observatory mission or even by the forthcoming AGILE
Italian satellite (Mereghetti et al. \cite{mere01}).

In the case of early-type binaries the situation is different. In
a typical WR+OB binary the collision of the winds from both stars
produces a strong shock at some point between the stars, in a
region exposed to strong UV stellar fields. Both electrons and
protons can be accelerated in these colliding wind regions
(Eichler \& Usov \cite{eichler93}). Electrons will cool through
synchrotron and inverse Compton (IC) radiation. In fact, the
detection of non-thermal radio emission in many early-type
binaries corroborates the existence of a population of
relativistic electrons in some of these systems (e.g. Dougherty \&
Williams \cite{dougwi00} and references therein). In some cases,
like Cygnus OB2 No. 5, the colliding wind region is spatially
resolved with VLA observations and it appears as an extended,
lightly elongated non-thermal feature in the radio images
(Contreras et al. \cite{contre97}). In such cases, the geometry of
the system can be established, radio flux components can be
separated, and detailed calculations of the expected IC emission
at high energies can be done. For the particular case of Cyg OB2
No. 5, for instance, Benaglia et al. (\cite{benaglia01}) estimate
that about a half of the gamma-ray flux measured by EGRET from the
Cyg OB2 association could come from the early-type binary.

Using data from the third and final EGRET catalog of point-like
gamma-ray sources (Hartman et al. \cite{hartman99}), a correlation
analysis between unidentified sources on the one hand and WR stars
on the other shows several coincidences (Romero et al.
\cite{romeros99}). The a priori chance probability is estimated in
the range $10^{-2}-10^{-3}$, which is suggestive, but certainly
not overwhelming as for the case of supernova remnants (Romero et
al. \cite{romeros99}, Romero \cite{romero01}). Some positional
associations, however, deserve further study in the light of the
most recent observations. This is particularly true for WR 140, a
very interesting system whose possible high-energy emission has
been already discussed in the pre-EGRET era by Pollock
(\cite{pollock87}) and also by Eichler \& Usov (\cite{eichler93})
in their seminal paper on non-thermal radiation from WR+OB
binaries. The location of WR 140 is within the 98\% confidence
contour of the gamma-ray source 3EG J2022+4317. The existence of
recent and detailed radio observations of this WR binary provides
new tools for a reliable estimate of the expected gamma-ray flux,
which can be now compared with the measured EGRET flux, not
available at the time of Eichler \& Usov's paper.

Physical conditions in WR binary systems with non-thermal
colliding wind regions can be probed through gamma-ray
observations. When thermal and synchrotron components of the radio
emission can be adequately separated, the orbital parameters are
well known, and the spectral types of the stars are determined,
then the ratio of the synchrotron to the gamma-ray luminosity can
be used to estimate the magnitude of the magnetic field in the
shocked region. Even in case of a lack of clear gamma-ray
detection, due to the instrument sensitivity, we can set bounds on
the magnetic field strength, and hence make inferences on the
field in the stars. The key point is that, granted the presence of
a population of relativistic particles and an UV-photon field,
both of which we know are present in WR binaries with non-thermal
radio emission, then the production of IC gamma-rays is
unavoidable. The question, of course, is whether these gamma-rays
come with a flux density high enough as to be detected by the
current technology.

In this paper we will study the case of three WR binaries: WR 140,
WR 146, and WR 147. We will use the most recent results from radio
observations in order to fix the value of model parameters, and
then we will compute the expected gamma-ray luminosity with
reasonable assumptions for the magnetic fields. We then will
compare with gamma-ray observations in order to test our original
assumptions. Finally, within the constrains imposed by the
observations, we will make some predictions for future gamma-ray
instruments like INTEGRAL and GLAST.

We will not discuss cases of very close (i.e. short period) binary
systems because they have typical separations of $\sim 0.1$ AU,
which impose an upper limit to the size of the particle
acceleration region at the colliding wind shocks that results, in
turn, in a severe constraint for the highest possible energy of
the relativistic electrons. Even with strong magnetic fields of
$\sim 1$ Gauss in the shocked region, electrons could not go well
within the GeV domain. However, soft gamma-rays of a few MeV are
still possible, and some of these systems should deserve further
high-energy studies as potential targets for the INTEGRAL
satellite, which is optimized for such an energy range.

The structure of the paper is as follows. In the next section we
will present the model for the gamma-ray emission from the
colliding wind region. In Section \ref{candidates} we will discuss
the main characteristics of the WR binaries in our sample. Section
\ref{gamma-observs} presents the existing gamma-ray information
about the regions where the stars are located. In Section
\ref{results} we give the main results whereas in Section
\ref{discussion} we discuss some of their implications. We close
in Section \ref{concls} with the conclusions.

\section{Gamma-ray emission from massive binaries}
\label{gamma-emission}

Gamma-ray production in early-type binaries with colliding winds
has been discussed by Eichler \& Usov (\cite{eichler93}), White \&
Chen (\cite{whichen95}), Benaglia et al. (\cite{benaglia01}) and
M\"ucke \& Pohl (\cite{mucke02}). We will present here the main
features of the basic model.

In an early-type binary system the winds from the primary (e.g. a
WR) and the secondary (e.g. an OB) stars flow nearly radially and
collide at a point located at a distance $r_{i}$ from the
$i$-star, given by :

\begin{equation}
r_1=\frac{1}{1+\eta^{1/2}}D,\;\;\;\;\;r_2=\frac{\eta^{1/2}}{1+\eta^{1/2}}D.
\end{equation}
In these expressions the subscript ``1'' stands for the primary
star, and ``2'' for the secondary, $D$ is the binary separation,
and the parameter $\eta$ is defined in terms of the wind terminal
velocities $v_{\infty}$ and the stellar mass loss rates $\dot{M}$:
\begin{equation}
\eta=\frac{\dot{M_2}v_{\infty,2}}{\dot{M_1}v_{\infty,1}}.\label{eta}
\end{equation}

At the colliding wind region, first-order diffusive shock
acceleration (e.g. Drury \cite{drury83}) results in the production
of a power-law spectrum $N(\gamma)\propto\gamma^{-p}$, with
$p\simeq2$ for electrons of energy $E=\gamma m_{\rm e} c^{2}$.
Because of the strong photon fields existing in the colliding wind
zone, IC losses are expected to dominate the radiative cooling at
high energies. The high energy cutoff of the electron spectrum can
be then obtained from the condition that the IC loss rate for the
particles does not exceed the acceleration rate (Eichler \& Usov
\cite{eichler93}):

\begin{eqnarray}
\gamma_{\rm max}^2 &\simeq& 3 \times 10^8 \eta \left(
\frac{v_{\infty, 1}}{2 \times 10^8\, {\rm cm}\,{\rm s}^{-1}}
\right)^2\, \left(\frac{B}{\rm G} \right) \, \times \nonumber
\\& & \left( \frac{D}{ 10^{13}\, {\rm cm}} \right)^2 \, \left(
\frac{L_{\rm OB}}{10^{39}\,{\rm erg}\,{\rm s}^{-1}} \right)^{-1}.
\label{g-max}
\end{eqnarray}

Here $B$ is the local value of the magnetic field and $L_{\rm OB}$
is the bolometric luminosity of the OB star, which is closer to
the shock. Typical values for the maximum Lorentz factor are,
according to Eq. (\ref{g-max}), of a few times $10^{4}$.

Inverse Compton losses will also produce a modification in the
spectrum of the relativistic particles. A break is expected at the
energy at which the cooling and escape times are equal (e.g.
Longair \cite{longair97}, p. 281). This will occur at a Lorentz
factor $\gamma_{\rm b}$ given by:

\begin{equation}
\gamma_{\rm b}=\frac{3m_e c^2}{4\sigma_{\rm T} U t_{\rm esc}},
\end{equation}

\noindent where $U$ is the average energy density of the photon
field, $t_{\rm esc}$ is average time spent by the particles in the
field region (typically $t_{\rm esc}\sim s/c$, with $s$ the linear
size of the colliding wind region), $m_{\rm e}$ is the electron
mass, and $\sigma_{\rm T}$ the Thomson cross section. The spectrum
will steepen from an index $p$ to $p+1$ for energies higher than
$\gamma_{\rm b}$.

The IC photons produced in a stellar photon field with seed
photons of frequency $\nu_{*} = 5.8\; 10^{10} \,\, T_{\rm eff}$,
where $T_{\rm eff}$ is the stellar effective temperature, are
characterized by $\nu^{\rm IC}=4/3 \gamma^{2} \nu_{*}$. The
electrons will also lose energy through synchrotron emission in
the magnetic field, producing a synchrotron flux with a spectrum
$S(\nu)\propto\nu^{-\alpha}$, with $\alpha=(p-1)/2$. The frequency
of the synchrotron photons will be $\nu_{\rm syn}= 4.2 \,\, B
\,\,\gamma^2 \,\, {\rm MHz}$. Depending on the strength of the
magnetic field, the tail of the synchrotron emission can reach
even optical frequencies in some cases. Of course, in such bands
it is not observable because of the overwhelming thermal
contribution, but at radio wavelengths it can be detected and
measured with interferometric instruments.

The rate of IC interactions in the colliding wind region per final
photon energy is:

\begin{eqnarray}
    \frac{dN}{dtd{\epsilon}}=\frac{3\sigma_{\rm T}c}{4\epsilon_{0}\gamma^{2}}f(x),
    \label{rate}
\end{eqnarray}

\noindent where $\epsilon_{0}$ is the typical energy of the seed
photons. For the case of Thomson scattering and adopting the
``head-on" approximation (i.e. the photons-- an isotropic field in
the lab frame-- are treated as coming from the opposite direction
to the electron's velocity in the electron frame, see Jones
\cite{jones68} for details), the function $f(x)$ can be written
as:

\begin{eqnarray}
    f(x)=\frac{2}{3}(1-x)P(1/4\gamma^{2},1,x),\ \ \
    x=\frac{\epsilon}{4\epsilon_{0}\gamma^{2}}.
\end{eqnarray}

The specific luminosity is then obtained by integrating the
scattering rate in Eq. (\ref{rate}) over the particle energy
distribution, and multiplying by the observed photon energy
$\epsilon$ and the number density $n_{\rm ph}=U/\epsilon_{0}$.
This leads to the expression:

\begin{eqnarray}    L_{\epsilon}&=&\frac{dL}{d{\epsilon}d\Omega}=
\frac{kV\sigma_{\rm T}m_{\rm e} c^3\:U}{8\pi\epsilon_{0}}
\frac{\epsilon}{\epsilon_{0}}\times\nonumber\\
&&\left\{(\gamma_{2})^{-(1+p)}\left(\frac{\epsilon}{4\epsilon_{0}
(3+p)(\gamma_{2})^{2}}-\frac{1}{1+p}\right)\right.\\
&&+\left.\left(\frac{\epsilon}{4\epsilon_{0}}\right)^{-(1+p)/2}
\frac{2}{(1+p)(3+p)}\right\}\label{lumtot}\nonumber
\end{eqnarray}

Here, $k$ is the constant involved in the energy distribution of
the electrons, $\gamma_2$ the highest energy of the electrons
considered in the calculation, and $V$ the interacting volume.
According to the energy range we are interested in, $\gamma_{\rm
2}$ can be equal to (or smaller than) $\gamma_{\rm max}$. When $p$
is steeper than 1 (as in our case) and $\gamma_{\rm
min}\ll\gamma_{\rm 2}$, this expression reduces to:

\begin{eqnarray}
    L_{\epsilon}=\frac{dL}{d{\epsilon}d\Omega}\approx \frac{kV\sigma_{\rm T}
    m_{\rm e} c^3\:U2^{p-1}}{\pi\epsilon_{0}(1+p)(3+p)}\left(\frac{\epsilon}
    {\epsilon_{0}}\right)^{-(p-1)/2}.
    \label{lumred}
\end{eqnarray}

A direct calculation of the IC luminosity of the colliding wind
region of a particular binary system requires a knowledge of the
parameter $k$ in the electron energy distribution. This parameter
is unknown, but we can use information on the synchrotron
luminosity of these same electrons in order to circumvent this
problem. In particular, the ratio of synchrotron to IC
luminosities for the colliding wind region can be estimated as
(White \& Chen \cite{whichen95}, Benaglia et al.
\cite{benaglia01}):

\begin{equation}
 \frac{L_{\rm
syn}}{L_{\rm IC}}=840 \frac{B^2 r_i^2}{L_i}, \label{LIC}
\end{equation}
where $B$ is expressed in G, $r_i$ is the distance to the $i$-star
in AU, and $L_i$ is the star luminosity in $L_{\sun}$ units. Since
$L_{\rm syn}$, $r_i$, and $L_i$ can all be, at least in principle,
determined from observations, the gamma-ray visibility of the
massive binary will crucially depend on the value of the magnetic
field in the particle acceleration region.

It is usually assumed that the external magnetic field of the star
in the absence of stellar wind is dipolar, and that with wind it
obeys the standard $B\propto r^{-1}$ radial dependence for large
$r$ given by Eichler \& Usov (\cite{eichler93}):

\begin{equation}
B\approx B_{*} \frac{V_{\rm
rot}}{v_{\infty}}\frac{R_{*}^{2}}{r_{\rm A }r}\;\;{\rm valid\;
for}\; r>R_{*}\frac{v_{\infty}}{V_{\rm rot}},
\label{B}\end{equation}

\noindent where $R_{*}$ is the star radius, $B_{*}$ the surface
magnetic field, $V_{\rm rot}$ the surface rotational velocity, and
$r_{\rm A}\approx R_{*}$ the Alfv\'en radius. The value of the
surface magnetic field is not well known, but it is thought that
in WR stars could reach $10^4$ G (e.g. Ignace et al.
\cite{ignace98}). Then, in the colliding wind region of typical
WR+OB binaries, the field could be in the range $10^{-2}-10$ G.

At the colliding wind zone an equipartition magnetic field can be
derived following Miley (\cite{miley80}):

\begin{eqnarray}
B_{\rm eq} &= \left[ \frac{2.84\;10^{-4}}{\theta_x\, \theta_y\,
s\, (\sin
\phi)^{3/2}}\;\frac{1+\chi}{f}\;\frac{F_o}{\nu_o^\alpha}\; \frac
{\nu_2^{\alpha + 1/2} - \nu_1^{\alpha + 1/2}}{\alpha + 1/2}
\right] ^ {2/7}\;{\rm G.} \label{B-Miley}
\end{eqnarray}

Here $\theta_x$, $\theta_y$ represent the source sizes in mas;
$\chi$ is the energy ratio between heavy particles and electrons;
$f$ is the filling factor of the emitting region; $s$ is the path
length through the source in the line of sight, in AU; $\phi$ is
the angle between the uniform magnetic field and the line of
sight; $F_0$ the flux density in mJy of the region at frequency
$\nu_0$; and $\nu_1$  and $\nu_2$ are the upper and lower cut off
frequencies presumed for the radio spectrum, these last three in
GHz. For minimum energy conditions, $\chi = f = \sin \phi = 1$.

IC scattering of stellar photons is not the only mechanism capable
of producing gamma-rays in the colliding wind region of an
early-type binary. Relativistic bremsstrahlung in the ions of the
winds will also result into gamma-ray emission. On the other hand,
the same first-order acceleration mechanism that forms the
electronic relativistic population should operate upon protons and
ions. The decay of neutral pions generated in the hadronic
interactions between the relativistic protons and the nuclei in
the wind yields gamma-rays, which at high energies present the
same energy spectrum than the parent proton population. At 67.5
MeV the spectrum should present the typical pion bump. All these
complementary contributions, however, are rather minor in
comparison with the IC emission because of the ambient densities
involved in colliding wind binaries. In Section \ref{results},
when we will present the results of our calculations of the IC
gamma-ray luminosity for several systems, we will also provide the
expected bremsstrahlung and hadronic gamma-ray luminosities for
completeness, but the reader is referred to Benaglia et al.
(\cite{benaglia01}) and references therein for the formulae.

\section{Selected binaries} \label{candidates}

The recently updated catalogue of WR stars (van der Hucht
\cite{hucht01}) lists about 230 objects, all but one at distances
up to 20 kpc. Two main radio surveys with high-resolution have
been carried out. The first one, using the VLA (Abbott et al.
\cite{abbott86}) compiled data from $\sim$ 40 stars up to 3 kpc,
north of declination $-47^0$. The southern one was done using ATCA
(Leitherer et al. \cite{leitherer97}, Chapman et al.
\cite{chapman99}) and encompasses $\sim$ 40 WR stars with $\delta
< 0^0$ and distances up to 3 kpc. Special objects were re-observed
in recent years (e.g. Contreras et al. \cite{contre97}, Dougherty
et al. \cite{dough96}, \cite{dough00}, etc). Some objects show
spectral indices $\alpha$ approaching the canonical thermal value
of $+0.6$, whereas the rest present negative and/or variable
values.

For the present investigation, we looked first for candidates with
detected non-thermal emission, indicating the presence of
relativistic electrons. Taking into account the works mentioned
above, a list can be made containing the objects that have shown
spectral indices other than thermal. Such a list is presented in
Table~\ref{table1}, where we have separated the stars in three
groups or types, according to the spectral index: `Comb' sources
are those that present an index near 0 or variable; `NT' ones have
a clearly negative value of $\alpha$. Two systems - WR 146 and WR
147- have been `{\sc resolved}' using VLA and MERLIN in two
sources each: one thermal, the other not.  The spectral
classification, distances, and periods in the table are taken
mostly from van der Hucht (\cite{hucht01}). Special references are
given for the spectral indices. We also provide information on the
different wavelengths at which the stars were detected, whenever
available.

After eliminating objects with no evidence of colliding winds or
close binaries, we are left with WR 146 and WR 147 as the best
candidates. The colliding wind region has been resolved in
continuum radio observations for these systems. They have orbital
periods of hundreds of years. The geometry of the colliding wind
region is relatively clear for these binaries.

A third classical candidate was added to our sample: WR 140.
Although the colliding wind region is not well-resolved here, this
star has been monitored over its complete period, it has been
pointed as the probable counterpart of an EGRET gamma-ray point
source (Romero et al. \cite{romeros99}), and it was also the
original example discussed by Eichler \& Usov (\cite{eichler93})
to illustrate their non-thermal model.

\begin{table*}
\caption[]{Wolf-Rayet stars with non-thermal -or combined-
emission.}
\begin{center}
\begin{tabular}{r c c c r c c }
&&&&&&\\
 \hline

WR &  Sp. Class.$^{\rm (a)}$ & $d$ & Period$^{\rm (a)}$ &
$\alpha^{(*)}$ & Type & Observations \\
  &    &   (kpc) & (days) & & &   (cm) \\  \hline &&&&&&\\

11 &  WC8 + O7.5III-V & 0.26$^{\rm (a)}$ & 78.53 & [$-0.5$]$^{\rm
(b)}$ & Comb& 3,6,13,20 \\ &&&&&&\\

14 &  WC7 + ? & 2.00$^{\rm (a)}$  &  2.42  & $-1.0^{\rm (b)}$ & NT
& 3,6
\\ &&&&&&\\

39 &  WC7 + OB?  &  5.53$^{\rm (a)}$   & --- & 0.0$^{\rm (b)}$  &
Comb  & 3,6,13,20
\\ &&&&&&\\

48 &  WC6(+O9.5/B0Iab) & 2.27$^{\rm (a)}$ &18.34 & $-0.4$,
$[-0.8]^{\rm (b)}$ & NT & 3,6,13\\ &&&&&&\\

90 &  WC7        &   1.64$^{\rm (a)}$    & --- & 0.0$^{\rm (b)}$ &
Comb & 3,6,13\\ &&&&&&\\

105&  WN9h        &    1.58$^{\rm (a)}$   & --- & $-0.3^{\rm (b)}$
& Comb & 3,6 \\ &&&&&&\\

112 & WC9d + OB ?   &4.15$^{\rm (a)}$      &  ---     &
$-1.35^{\rm (b)}$ & NT & 3,13\\ &&&&&&\\

125 & WC7ed + O9III   & 3.06$^{\rm (a)}$ & $>$6600  &
$-0.5\rightarrow 0.7^{\rm (c),(d)}$  &Comb& 2,6,20 \\ &&&&&&\\

137 & WC7pd + O9  & 2.38$^{\rm (a)}$ &  4765& 0.0$^{\rm (e)}$ &
Comb & ---\\ &&&&&&\\

140 & WC7pd + O4-5 &  1.10$^{\rm (a)}$,$1.3^{\rm (f)}$ & 2900
&$-0.5^{\rm (g)}$,$-0.6^{\rm (h)}$& Comb& 2,6,20 \\ &&&&&&\\

146 &  WC6 + O8  &  0.72$^{\rm (a)}$,1.25$^{\rm (h)}$ & ELPB$^{\rm
(i)}$ &$-0.62^{\rm (j)}$ & {\sc resolved}&1.3,3,6,20 \\ & &&&&&\\

147 & WN8(h) + B0.5V & 0.65$^{\rm (a)}$ & ELPB$^{\rm (i)}$ &
$-0.43^{\rm (k)}$,$-1.1^{\rm (l)}$ & {\sc resolved} &
1.3,2,3.6,20\\

&&&&&&\\
 \hline

\multicolumn{7}{l} {(*) Brackets mean non-thermal index derived
from model; (a) van der Hucht \cite{hucht01}; (b) Chapman et al.
\cite{chapman99}; } \cr

\multicolumn{7}{l} {(c) Abbott et al. \cite{abbott86}; (d)
Williams et al. \cite{wills92}; (e) Dougherty \& Williams
\cite{dougwi00}; (f) Smith et al. \cite{smith90}; }\cr

\multicolumn{7}{l} {(g) White \& Becker \cite{whibec95}; (h) van
der Hucht et al. \cite{huchts01}; (i) stands for Extremely Long
Period Binary;}\cr

\multicolumn{7}{l} {(j) Dougherty et al. \cite{dough00}; (k) Setia
Gunawan et al. \cite{setia01a}; (l) Skinner et al.
\cite{skinner99}}\cr

\end{tabular}
\end{center}
\label{table1}
\end{table*}

\subsection{WR 140}

WR 140 (HD 193793, V1687 Cyg) is a spectroscopic binary system,
formed by a WC7pd and an O4-5 companion (van der Hucht
\cite{hucht01}). It was the first star for which non-thermal
emission has been detected (Florkowski \& Gottestman
\cite{flork77}). The condition of colliding wind binary was
confirmed by Williams et al. (\cite{wills87}), deducing a period
of 7.9 yr and an eccentricity $e=0.84$. It has shown variability
at radio, IR, optical, UV, and X-rays (Setia Gunawan et al.
\cite{setia01b} and references therein). Periodic dust formation
around periastron passage would be responsible for the IR excess
observed (Williams et al. \cite{wills87}).

Both White \& Becker (\cite{whibec95}) and Williams et al.
(\cite{wills90}, \cite{wills94}) monitored the system at radio
frequencies. The first authors used the VLA at 2, 6 and 20 cm,
from 1985 to 1993. Williams et al. observed WR 140 from 1986 to
1994 with WSRT. The observations allow to estimate a spectral
index of $-0.6$ (van der Hucht et al. \cite{huchts01}). They
adopted a model in which the WR stellar wind is strongly enhanced
at the equatorial plane, where most of the mass loss is confined.
The emission suffers from varying circumstellar extinction in the
line of sight to the non-thermal radio source that is identified
with the colliding wind region. The orbit should be inclined. More
recently, Setia Gunawan et al. (\cite{setia01b}) presented the
results of 16 years of UV monitoring of WR 140, and provided
explanations for the observed spectral variability.

\begin{table*}
\caption[]{Stellar parameters adopted for the WR 140 system}
\begin{center}
\begin{tabular}{l r r r l}
 \hline
 Variable & WR& & OB & unit \\

 \hline

Separation from colliding wind region & 8.5$^{\rm (*)}$ &&
1.5$^{\rm (*)}$ & AU \cr

Spectral Class. & WC7pd$^{\rm (a)}$ && O4-5 (V)$^{\rm (b)}$ & \cr

log ($L/{\rm L}_\odot$) & 5.35$^{\rm (c)}$ && 5.80$^{\rm (d)}$ \cr

$T_{\rm eff}$ & 48000$^{\rm (c)}$ && 47400$^{\rm (d)}$ & K \cr

$R_*$ & 6.0$^{\rm (c)}$ && 12.0$^{\rm (d)}$ & R$_\odot$ \cr

$\mu$ & 5$^{\rm (c)}$ && 1.5$^{\rm (c)}$ & \cr

$v_\infty$ & 2860$^{\rm (e)}$ && 3100$^{\rm (e)}$ & km s$^{-1}$
\cr

$\dot{M}$ & 5.7 10$^{-5}$$^{\rm (e)}$ && 3.0 10$^{-6}$$^{\rm (c)}$
& M$_\odot$ yr$^{-1}$ \cr

\hline

Distance $d$&& 1320$^{\rm (e)}$ & & pc \cr

$\eta$ && 0.06$\,\,\,\,\,\,$ &&   \cr

Non-thermal spectral index && $-0.60$$^{\rm (e)}$ & & \cr

$S_{5 \rm GHz}$ && 22.5$^{\rm (e)}$ & & mJy \cr

Size of non-thermal source && 6.0 10$^{13}$$^{\rm (c)}$ & & cm \cr

Photon energy density $U$ && 38.5$\,\,\,\,\,\,$ && erg
cm$^{-3}$\cr

$B_{\rm cwr}$ (equipartition) && 200$^{\rm (c)}$ & & mG \cr

\hline \multicolumn{5}{l} {(*) for binary separation of 10 AU; (a)
van der Hucht \cite{hucht01}; (b) Setia Gunawan et al.
\cite{setia01b}; }\cr \multicolumn{5}{l} {(c) adopted here, see
text; (d) Vacca et al. \cite{vacca96}; (e) van der Hucht et al.
\cite{huchts01}}\cr

\end{tabular}
\end{center}
\label{table2}
\end{table*}

The stellar parameters of WR 140 relevant to this work are listed
in Table~\ref{table2} along with the corresponding references.
Throughout the period, the separation between the stellar
components of the system ranges from 2.4 to 27 AU. We have assumed
a separation $D = 10 $ AU in order to compare with gamma-ray
observations (see Section \ref{gamma-observs}), and computed a
value $\eta = 0.06$, that was used in our calculations (Section
\ref{results}). The WR stellar luminosity was taken as an average
of all model luminosities given by Koesterke \& Hamann
(\cite{koestha95}) for WC7 stars. The luminosity class of the
secondary is suggested by Setia Gunawan et al. (\cite{setia01b})
as O4-5(V). The effective temperature and radius of the primary
were assumed the same of WR 125 (Koesterke \& Hamann
\cite{koestha95}), because of their similarities. The effective
temperature, stellar luminosity, and radius for the secondary were
taken from the tables of Vacca et al. (\cite{vacca96}), for a
spectral type O4.5~V. A mean molecular weight ($\mu$) of 5 was
adopted for the primary, and 1.5 for the secondary. The mass loss
rate of the secondary was computed following the models of Vink et
al. (\cite{vink00}).

Van der Hucht et al. (\cite{huchts01}) give values for mass loss
rates and wind terminal velocities of both components, the
spectral non-thermal index, distance, and non-thermal flux density
at 6 cm. In order to obtain an estimate of the extension of the
colliding wind region, we have computed the radii of the
photospheres at 5 GHz of both components, using the expression
from Wright \& Barlow (\cite{wribar75}). After verifying they
overlap, we have considered the size of the non-thermal source
equal to the OB radio photosphere, about 4 AU.

The equipartition magnetic field can be calculated with Eq
(\ref{B-Miley}) using $\nu_1 = 1.4$ GHz and $\nu_2 = 250$ GHz (see
Wendker \cite{wendker95}), and assuming a spherical non-thermal
source with $\theta_x = \theta_y = 3$ mas, and $s = 4$ AU.
Inaccuracies in the colliding wind region size or in the filling
factor introduce the main errors in the determination of the
magnetic field. For example, if $f=0.1$, or $s=2$ AU, the magnetic
field doubles its value.







\subsection{WR 146}

This system contains the brightest WR star detected in radio
continuum so far. The secondary is an OB star in an orbit with a
period estimated in $\sim$ 300 yr (Dougherty et al.
\cite{dough96}).

After observing with the EVN and VLA, Felli \& Massi
(\cite{felli91}) detected emission with a spectral index $\alpha =
-1$. Dougherty et al. (\cite{dough96}) resolved the system using
MERLIN in two sources: a bright non-thermal northern component of
$38\pm 1$ mas in diameter, and a weaker southern component.
Niemela et al. (\cite{niemela98}) reported the resolution of the
system after HST-{\sc wfpc2} observations. Their result for the
binary separation was confirmed by 22 GHz observations taken by
Dougherty et al. (\cite{dough00}), who detected the stellar wind
of the companion star with the VLA. Their radio maps allowed to
derive a value for $r_2$ of $\sim 50$ AU, at $d = 1.25$ kpc, and
$\eta = 0.1$. They took optical spectra of the system and
interpreted the results as strongly suggesting that the O8 star is
an Of, possibly a supergiant.

Setia Gunawan et al. (\cite{setia00}) presented the results of 1.4
and 5 GHz observations with WSRT, from 1989 to 1999. They found
three different kinds of flux variability behaviour on this
time-span: a linear increasing trend observed during 10 years, a
possible 3.38-yr periodic signal, and rapid fluctuations on time
scales of weeks; they also derived a time-averaged non-thermal
spectral index $\alpha = -0.62$.

\begin{table*}[]
\caption[]{Stellar parameters adopted for the WR 146 system}
\begin{center}
\begin{tabular}{l r r r l}
 \hline
Variable & WR & & OB & unit \\

 \hline

Separation from colliding wind region & 160$^{(*)}$ && 50$^{(*)}$
& AU \cr

Spectral Class. & WC6$^{\rm (a)}$ && O8 If$^{\rm (b)}$ & \cr

log ($L/{\rm L}_\odot$) & 5.10$^{\rm (c)}$ && 5.00$^{\rm (d)}$ \cr

$T_{\rm eff}$ & 49000$^{\rm (e)}$ && 35700$^{\rm (f)}$ & K \cr

$R_*$ & 5.0$^{\rm (e)}$ && 23.1$^{\rm (f)}$ & R$_\odot$ \cr

$\mu$ & 5.29$^{\rm (g)}$ && 1.34$^{\rm (h)}$ & \cr

$v_\infty$ & 2700$^{\rm (g)}$ && 1300$^{\rm (i)}$ & km s$^{-1}$
\cr

$\dot{M}$ & 2.6 10$^{-5}$$^{\rm (g)}$ && 5.4 10$^{-6}$$^{\rm (d)}$
& M$_\odot$ yr$^{-1}$ \cr

\hline

Distance $d$&& 1250$^{\rm (i)}$ & & pc \cr

$\eta$ &&  0.10$\,\,\,\,\,\,$&&   \cr

Non-thermal spectral index && $-0.62$$^{\rm (c)}$ & & \cr

$S_{5 \rm GHz}$ && 28.5$^{\rm (i)}$ & & mJy \cr

Size of non-thermal source && 7.0 10$^{14}\,^{\rm (d)}$ & & cm \cr

Photon energy density $U$ && 6.1\,10$^{-3}$$\,\,\,\,\,\,$ && erg
cm$^{-3}$\cr

$B_{\rm cwr}$ (equipartition) && 25.0$^{\rm (d)}$ & & mG \cr

\hline \multicolumn{5}{l} {(*) deduced for $r_2 = 40\pm9$ mas
(Dougherty et al. \cite{dough00}); (a) Smith et al.
\cite{smith90};}\cr

\multicolumn{5}{l} {(b) Dougherty et al. \cite{dough00}; (c) Setia
Gunawan et al. \cite{setia00}; (d) adopted here, see text;}\cr

\multicolumn{5}{l} {(e) Koesterke \& Hamann \cite{koestha95}; (f)
Vacca et al. \cite{vacca96}; (g) Willis et al. \cite{willis97};
}\cr

\multicolumn{5}{l} {(h) Lamers \& Leitherer \cite{lamleith93}; (i)
van der Hucht et al. \cite{huchts01}}
\end{tabular}
\end{center}
\label{table3}
\end{table*}

The values adopted for our calculations are shown in Table
\ref{table3}. Since the distance estimates to the WR 146 binary
system differ from 0.75 kpc to 1.7 kpc (Setia Gunawan et al.
\cite{setia00} and references therein), we have carried out the
calculations adopting the mean value $d = 1.25$ kpc (van der Hucht
et al. \cite{huchts01}). The corresponding separation of the two
stars is 210 AU (Setia Gunawan et al. \cite{setia00}). The stellar
luminosity of the secondary is an average of the values given by
Setia Gunawan et al. (\cite{setia00}). The effective temperature
and radius of the secondary were taken from the tables of Vacca et
al. (\cite{vacca96}). The mass loss rate of the secondary was
computed from Eq. (\ref{eta}). The colliding wind region size was
derived assuming an extension of the non-thermal source of 46 AU
at 1.25 kpc (Setia Gunawan et al. \cite{setia00}). We estimated
the equipartition magnetic field between 327 MHz and 22 GHz
(Taylor et al. \cite{taylor96}), with $\theta_x = \theta_y = 38$
mas, and $s = 47$ AU.

\subsection{WR 147} \label{candi-wr147}

Non-thermal emission form this system was reported after VLA
observations by Abbott et al. (\cite{abbott86}). Higher angular
resolution observations taken with the same instrument (Churchwell
et al. \cite{church92}) and MERLIN 5-GHz observations by Moran et
al. (\cite{moran89}) confirmed the presence of two sources: a
southern thermal one (WR 147S) superposed with the WR star, and a
northern non-thermal component (WR 147N). Williams et al.
(\cite{wills97}) observed the system again with MERLIN and found
that the two sources are separated by $575\pm 15$ mas. They
reported also a faint IR source near WR 147N, but slightly farther
to the WR star than the radio source WR 147N, and derived a
spectral type B0.5 V for the IR star (confirmed later through
optical observations by Niemela et al. \cite{niemela98}). Because
of the presence of non-thermal emission between the two stars and
its location much closer to that with the weaker wind, Williams et
al. (\cite{wills97}) proposed the system as a colliding wind
binary.

Using VLA observations at 3.6 cm, Contreras \& Rodr\'{\i}guez
(\cite{contre99}) found that the non-thermal wind-interaction zone
remained constant in flux density during 1995-1996. The thermal
emission from the WR star, on the contrary, increased $\sim25$\%,
probably reflecting the inhomogeneous nature of the wind. Both the
observed radio morphology and the theoretical modeling by
Contreras \& Rodr\'{\i}guez (\cite{contre99}) clearly supports a
colliding winds scenario for WR 147.

Setia Gunawan et al. (\cite{setia01a}) presented the results of a
monitoring campaign of the source from 1988 to 1997 at 1.4 and 5
GHz with WSRT. Once subtracted the southern thermal contribution,
the spectral energy distribution could be fitted by a synchrotron
emission model which includes free-free absorption. Flux density
variations on different time scales can be explained by
considering inhomogeneities in the wind, plasma outflow, etc.
After fitting the thermal source data, they computed a non-thermal
flux density $S_{\rm NT} = S_{\rm total} - S_{\rm fit}$, obtaining
the spectral indices $\alpha_{\rm thermal} = +0.6$ and
$\alpha_{\rm syn} = -0.43$ for the thermal and synchrotron
components, respectively. The statistical significance of this
result, however, is very low. Skinner et al. (\cite{skinner99})
have concluded, using nearly simultaneous observations at five
frequencies, that there is a significant steepening in the
non-thermal spectrum above 10 GHz. These authors have even
suggested that a monoenergetic relativistic electron spectrum
injected at the source might be a plausible explanation of the
observed radio flux distribution. The quality of the radio data,
unfortunately, does not allow to draw any firm conclusion.


The parameters for the system WR 147 adopted in our calculations
are listed in Table \ref{table4}, with the respective references.
A canonical value of $\alpha = -0.5$ was adopted, corresponding to
the expected $p=2$ spectrum in the electron distribution at
relatively low energies. Compton losses will introduce a
steepening at higher energies. In Section \ref{discussion} we will
revisit this assumption and we will briefly discuss the
implications of the monoenergetic electron distribution suggested
by Skinner et al. (\cite{skinner99}) for the high-energy emission.
The equipartition field at the colliding wind region was computed
for $\theta_x = \theta_y = 267$ mas, and $s = 175$ kpc, between
frequencies 350 MHz (Setia Gunawan et al. \cite{setia01a}) and 250
GHz (Wendker \cite{wendker95}).

\begin{table*}[]
\caption[]{Stellar parameters adopted for the WR 147 system}
\begin{center}
\begin{tabular}{l r r r l}
 \hline
 Variable & WR & &OB & unit \\

 \hline

Separation from colliding wind region & 374$^{\rm (a)}$ &&
43$^{\rm (b)}$ & AU \cr

Spectral Class. & WN8(h)$^{\rm (c)}$ && B0.5 V$^{\rm (a)}$ & \cr

log ($L/{\rm L}_\odot$) & 5.52$^{\rm (d)}$ && 4.70$^{\rm (e)}$ \cr

$T_{\rm eff}$ & 26000$^{\rm (d)}$ && 28500$^{\rm (f)}$ & K \cr

$R_*$ & 20.6$^{\rm (e)}$& & 8.0$^{\rm (g)}$ & R$_\odot$ \cr

$\mu$ & 3.18$^{\rm (e)}$ && 1.5$^{\rm (b)}$ & \cr

$v_\infty$ & 950$^{\rm (e)}$ && 800$^{\rm (b)}$ & km s$^{-1}$ \cr

$\dot{M}$ & 2.4 10$^{-5}$$^{\rm (e)}$& & 4.0 10$^{-7}$$^{\rm (b)}$
& M$_\odot$ yr$^{-1}$ \cr

\hline

Distance $d$&& 650$^{\rm (e)}$ & & pc \cr

$\eta$ && 0.014$\,\,\,\,\,\,$&&   \cr

Non-thermal spectral index& & $-0.5$$^{\rm (i)}$ &   & \cr

$S_{5 \rm GHz}$ && 12.5$^{\rm (h)}$ & & mJy \cr

Size of non-thermal source && 2.6 10$^{15}$$^{\rm (b)}$ & & cm \cr

Photon energy density $U$ && 4.0\,10$^{-3}$\,\,\,\,\,\, && erg
cm$^{-3}$\cr

$B_{\rm cwr}$ (equipartition)& & 5.0$^{\rm (i)}$ & & mG\cr

\hline \multicolumn{5}{l} {(a) Williams et al. \cite{wills97}; (b)
Setia Gunawan et al. \cite{setia01a}, see text for a discussion;
(c) Smith}\cr

\multicolumn{5}{l} {et al. \cite{smith96}; (d) Crowther et al.
\cite{crowther95}; (e) Morris et al. \cite{morris00}; (f) Crowther
\cite{crowther97}; (g) Vacca et}\cr

\multicolumn{5}{l} {al. \cite{vacca96}; (h) van der Hucht et al.
\cite{huchts01}; (i) adopted here, see text}\cr
\end{tabular}
\end{center}
\label{table4}
\end{table*}

\section{Gamma-ray observations} \label{gamma-observs}

The gamma-ray observations used here were provided by the EGRET
telescope of the Compton satellite. The reduced data were
published in the third EGRET catalog (Hartman et al.
\cite{hartman99}). The catalog lists the best estimated position
of 271 point gamma-ray sources, along with the location error
boxes (taken as the 95\% confidence contours), integrated
gamma-ray fluxes in the energy range 100 MeV -- 20 GeV, spectral
indices, and other information.

\begin{figure}
\includegraphics[width=\hsize]{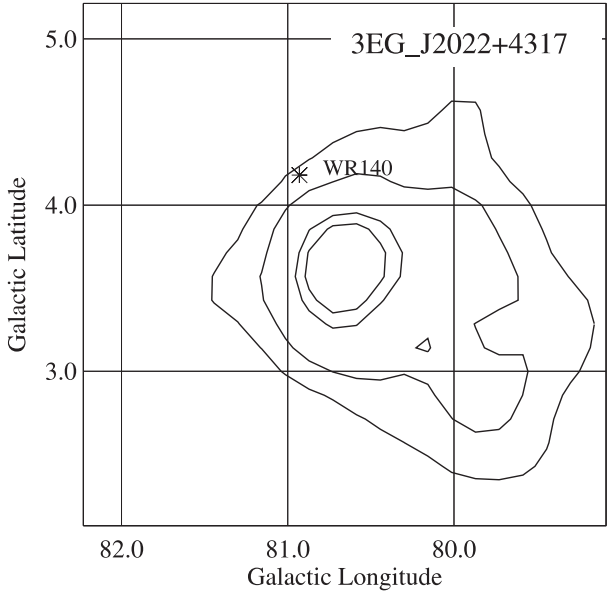}
\caption{EGRET probability contours for 3EG J2022+4317. Contour
labels are 50\% 68\%, 95\% and 98\%. The position of WR 140 is
marked} \label{fig1}
\end{figure}

Only WR 140 is positionally coincident with a point-like source in
the catalog, 3EG J2022+4317. The photon flux summed over satellite
observing cycles 1, 2, 3, and 4 is $24.7\pm5.2\,\, 10^{-8}$ ph
cm$^{-2}$ s$^{-1}$. The photon spectral index $F(E)\propto
E^{-\Gamma}$ is $\Gamma=2.31\pm 0.19$. The location error box is
large, with a radius of $\sim0.7$ degrees, which renders a clear
identification almost impossible in practice. However, WR~140 is
the only highly energetic source known in this region (Romero et
al. \cite{romeros99}). Figure~\ref{fig1}
 shows the relative position of WR 140 with respect to the
gamma-ray source.

The EGRET viewing periods for this source cover since 1991 to
1994. Following the orbital positions and times deduced by Setia
Gunawan et al. (\cite{setia01b}), the stars were separated between
$\sim$ 4 to $\sim$ 16 AU during the time-span 1991.5 - 1994.5. We
took a mean distance $D = 10$ AU for the binary separation during
the gamma-ray observations in order to perform the calculations of
the expected average flux.

\begin{figure*}
\begin{center}
\includegraphics[width=10cm]{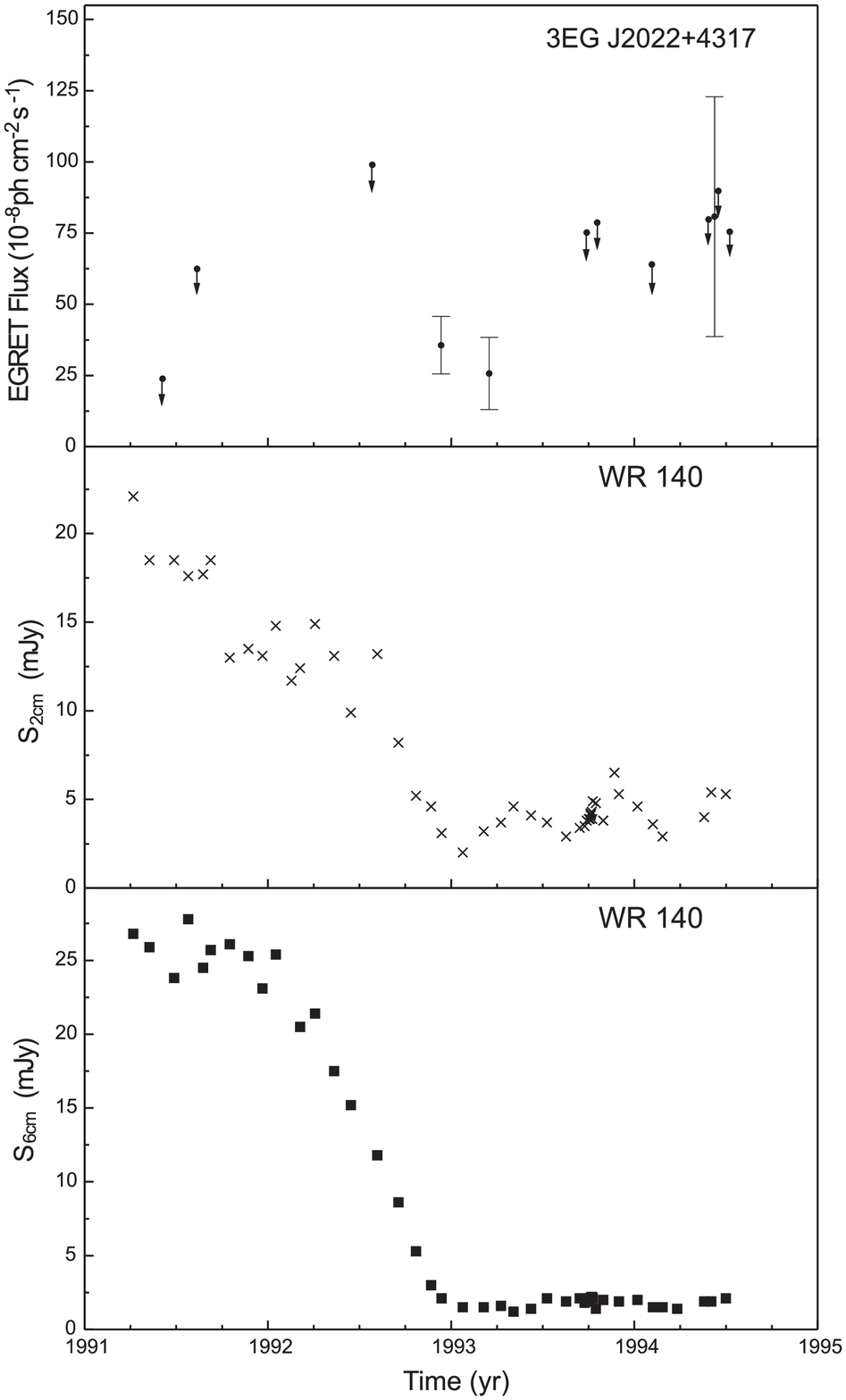}
\caption{Light curves at a) gamma-rays, b) 2cm, and c) 6cm.}
\label{fig2}
\end{center}
\end{figure*}

The variability analysis of the gamma-ray data for 3EG J2022+4317
gives a variability index $I = 1.3$ in Torres et al.
(\cite{torres01a}) scale, which is normalized to the average
pulsar variability. A source will be variable if $I
> 5$, at $8\sigma$ level. Non-variable sources have indices typically $I<1.7$
(Torres et al. \cite{torres01b}). Tompkins (\cite{tompkins99})
analysis yields for this source values of his $\tau$ index of $<
\tau> = 0.13$, $\tau_{\rm min} = 0.00$, and $\tau_{\rm max} =
0.50$. Here, a variable source will have $\tau_{\rm min} > 0.6$
(see Torres et al. \cite{torres01b} for a comparison of both
indices). Consequently, the source has not been formally variable
during the observing time interval. But a note of caution is
necessary here. As it can be seen in Fig.~\ref{fig2}, upper panel,
all but 3 observing periods for this source yielded only upper
limits. In such a case the formal variability index can be
misleading because under the upper limits real variability might
be hidden. In the lower panels of Fig.~\ref{fig2} we show the
radio evolution of the WR 140 for the same epoch of the gamma-ray
observations. The data is taken from White \& Becker
(\cite{whibec95}), directly from 1991 to 1993.6 and by
extrapolation from 1993.6 to 1994.5. We can see that the radio
flux was significantly decreasing from 1992 to 1993. If the value
of the gamma-ray flux of the source was close to the upper limit
established by EGRET in mid-1992, then the high-energy emission
might also have decreased significantly that year. The poor
resolution of the EGRET lightcurve prevents any conclusion in this
respect. It is not possible, in our opinion, to reject the
association of WR 140 with 3EG J2022+4317 on the sole basis on the
low variability indices.

Although the authors of the third EGRET catalog give the
approximate upper limits for gamma-ray sources at almost any point
in the sky, at the positions of WR 146 and WR 147 no threshold
could be computed. The reason is that the positions of the stars
are close to the fairly strong source 3EG J2033+4118. This makes
it impossible for the gamma reduction programmes to get a
statistically meaningful upper limit for weaker sources in the
surroundings. Because of this contamination problem, we consider
for the positions of WR~146 and WR 147, an EGRET detection limit
of 70\% of the flux of the gamma-ray source 3EG 2033+4118, i.e.
$50\; 10^{-8}$ ph cm$^{-2}$ s$^{-1}$, which seems to be a
reasonable assumption (Bob Hartman, private communication).

\section{Results} \label{results}

The following results for the different WR binaries in our sample
are obtained with the parameters listed in Tables
~\ref{table2}~-~\ref{table4}. A summary of the results is
presented in Table~\ref{table5}, where we list, for each system,
the expected gamma-ray luminosity due to inverse Compton
scattering, the relativistic bremsstrahlung, and the pion-decay
contributions to the gamma-ray flux generated at the colliding
wind region. In the following subsections we describe the main
results of our calculations for each star.

\subsection{WR 140}

For this system we have $B_{\rm cwr}\sim0.2$ G and a very high
photon energy density of $U=38.5$ erg s$^{-1}$. The inverse
Compton losses impose an upper energy limit for the locally
accelerated electrons of $\gamma_{\rm max} = 4.2 \; 10^{4}$ and a
break in the spectrum at $\gamma_{\rm b}\approx 400$. The break
will appear in the synchrotron spectrum at $\nu_{\rm b,\;
syn}\approx 135.2$ GHz and at gamma-ray energies at $E_{\rm
b}=2.45$ MeV, below the EGRET energy range. The Lorentz factors of
electrons radiating IC gamma-rays in the ambient photon fields and
contributing to 100 MeV$-$20 GeV are between $\gamma_{\rm 1} =
2.5\; 10^3$ and $\gamma_{\rm 2} = 3.6\; 10^4$. Electrons having
Lorentz factors in this range are also capable of emitting
synchrotron radiation between $\nu_{\rm 1} = 5.5\;10^3$ GHz and
$\nu_{\rm 2} = 1.1 \; 10^6$ GHz. The total synchrotron luminosity
from these electrons is $\sim 2.5\; 10^{30}$ erg s$^{-1}$.

The total gamma-ray luminosity due to IC scattering at the
colliding wind region between 100 MeV and 20 GeV results $\sim2.1
\; 10^{34}$ erg s$^{-1}$. At this energies, after the break in the
gamma-ray spectrum, the photon spectral index should be
$\Gamma\sim 2.1$ (original injection spectrum with $p=2.2$
according to radio observations below the break). Such an index is
in reasonable agreement with the index observed by EGRET in the
source 3EG J2022+4317 ($\Gamma=2.31\pm 0.19$), especially if we
take into account that there are also uncertainties in the
determination of the radio spectral index.

Both the luminosities due to pion decays from $p-p$ interactions
($< 10^{22}$ erg s$^{-1}$) and relativistic bremsstrahlung in the
winds ($< 10^{32}$ erg s$^{-1}$) can be disregarded in comparison
to the IC luminosity (see Benaglia et al. \cite{benaglia01} for
details of calculation).

If the unidentified gamma-ray source 3EG J2022+4317 is at the same
distance of WR 140, the measured EGRET flux implies a luminosity
of $\sim3.2\; 10^{34}$ erg s$^{-1}$, of the same order of
magnitude than the computed luminosity due to IC scattering at the
colliding wind region. Errors in the observed gamma-ray emission
are at the level of $\sim25$\%. An additional factor that can
affect the computations is the luminosity of the secondary star.
For the WR~140 system it is not derived directly from
observations, but interpolated from the tables of Vacca et al.
(\cite{vacca96}) and can be slightly overestimated. In any case,
it seems likely that significant part of the gamma-ray flux of 3EG
J2022+4317 might be contributed by the colliding wind region of
WR~140.

\begin{figure}
\begin{center}
\includegraphics[width=7cm]{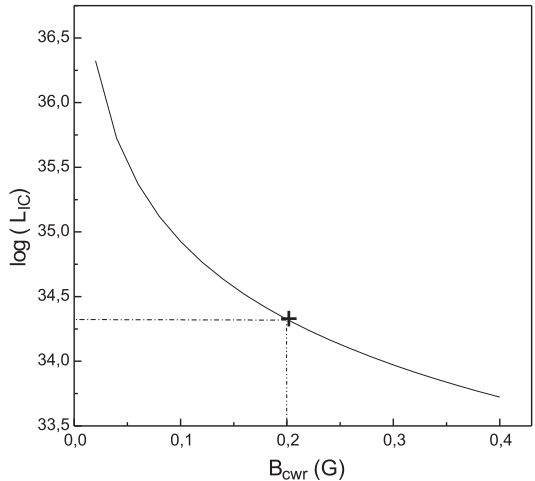}
\caption{Gamma-ray luminosity (in erg s$^{-1}$) between 100 MeV
and 20 GeV versus the magnetic field for WR 140. The horizontal
line indicates the expected luminosity of 3EG J2022+4317, if it is
located at the distance of WR 140.} \label{fig3}
\end{center}
\end{figure}

In Fig.~\ref{fig3} we present a plot of the gamma-ray luminosity
of the colliding wind region of WR 140 versus the assumed magnetic
field. The luminosity of 3EG J2022+4317 at the same distance is
indicated with a horizontal line. For high values of the magnetic
field the synchrotron losses dominate over IC losses, and the
gamma-ray emission is quenched. We emphasize, however, that in
order to explain the gamma-ray source 3EG J2022+4317 through WR
140, no extreme hypothesis is required, and the Occam's razor
principle is fulfilled since no new and otherwise yet undetected
object is postulated. In this sense, we suggest that WR 140 should
be considered as the best currently available explanation for the
origin of 3EG J2022+4317. The GLAST telescope, with its improved
source location accuracy, will be able to test this proposition.

\subsection{WR 146}

For this system the equipartition magnetic field is significantly
lower than in the case of WR 140: $B_{\rm cwr}\sim$ 25 mG. This
lower value is mainly due to the different size and location of
the non-thermal regions. With the values of the parameters listed
in Table~\ref{table3}, we obtain the following results.

The maximum energy of the accelerated particles in the colliding
wind region is $\gamma_{\rm max} = 6\;10^5$. These are very
energetic particles indeed. At such energies Klein-Nishina effects
are important. The highest energy for IC photons produced by these
electrons is $\sim \gamma_{\rm max}m_{\rm e} c^2 /2 \sim 0.2$ TeV.

The Lorentz factors of the electrons contributing to EGRET's
energy range are between $\gamma_{\rm 1} = 3\;10^3$ and
$\gamma_{\rm 2} = 4.2\;10^4$. The local energy density $U=6.1 \;
10^{-3}$ erg cm$^{-3}$ implies a break in the spectrum at very
high energies, within the optical region ($4.9\; 10^6$ GHz).
Electrons having Lorentz factors in the above range radiate
synchrotron photons between $\nu_{\rm 1} = 9.2\;10^2$ GHz and
$\nu_{\rm 2} = 1.8\;10^5$ GHz. The total synchrotron luminosity in
this frequency range is $3.3\;10^{31}$ erg s$^{-1}$.

The gamma-ray luminosity due to IC scattering at the colliding
wind region, contributed by seed photons from both stars is
$\sim2.8 \;10^{33}$ erg s$^{-1}$. In comparison, the gamma-ray
luminosity due to pion decay is $\sim2.5 \;10^{21}$ erg s$^{-1}$,
and to relativistic bremsstrahlung,  $\sim4.6\; 10^{29}$ erg
s$^{-1}$, resulting both negligible. The IC spectral break appears
at high energies (538 GeV), well above EGRET's range, so the
spectrum in the MeV-GeV band should be much harder than in the
case of WR 140, with values $\Gamma\sim 1.6$.

\begin{figure}
\begin{center}
\includegraphics[width=7cm]{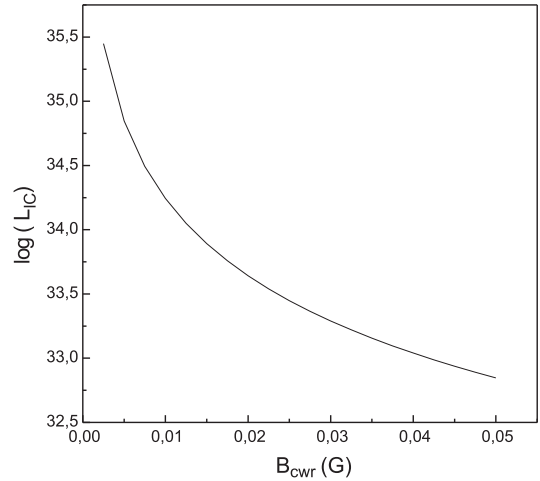}
\caption{Gamma-ray luminosity (in erg s$^{-1}$) between 100 MeV
and 20 GeV versus the magnetic field for WR 146.} \label{fig4}
\end{center}
\end{figure}

Figure~\ref{fig4} presents a plot of the gamma-ray luminosity between 100
MeV and 20 GeV versus the magnetic field. Very weak fields are
ruled out by the EGRET non-detection.

As we mentioned, the EGRET threshold towards the position of WR
146 can be taken as $50\,\,10^{-8}$ ph cm$^{-2}$ s$^{-1}$. This
means that a gamma-ray source at the distance of the binary system
would be detectable by EGRET if its luminosity is greater than
$8.0 \;10^{34}$ erg s$^{-1}$, which is more than an order of
magnitude above the expected luminosity coming from the colliding
wind region. This fact can explain why  no gamma-ray source was
detected towards WR 146. However, as we will see, the expected
flux is high enough as to be detected by GLAST, if the
contamination problem from nearby sources can be solved.

\subsection{WR 147}

For WR 147 we have  $B_{\rm cwr}\sim$ 5 mG. The maximum Lorentz
factor for the electrons imposed by the IC losses is $\gamma_{\rm
max} = 1.0\;10^5$. The high-energy cutoff frequency is then
$\nu_{\rm cutoff}= 2.1\; 10^{25}$ Hz or $\sim90$ GeV.

The Lorentz factors of the electrons whose IC emission falls in
EGRET's energy range are between $\gamma_{\rm 1} = 3.3\;10^3$ and
$\gamma_{\rm 2} = 4.7 \;10^4$. Their synchrotron radiation is
between $\nu_{\rm 1} = 230$ GHz and $\nu_{\rm 2} = 4.6\; 10^4$
GHz. The total synchrotron luminosity with such frequency interval
results $\sim5.7 \; 10^{30}$ erg s$^{-1}$. The break produced by
the local photon field of density $U=4.0 \; 10^{-3}$ erg cm$^{-3}$
in the electron spectrum occurs at a Lorentz  factor of
$\gamma_{\rm b}\approx 8.9\; 10^{4}$, and results in breaks in the
synchrotron (at $\sim 1.7 \; 10^{5}$ GHz) and IC spectrum (at
$\sim 72$ GeV).

\begin{figure}
\begin{center}
\includegraphics[width=7cm]{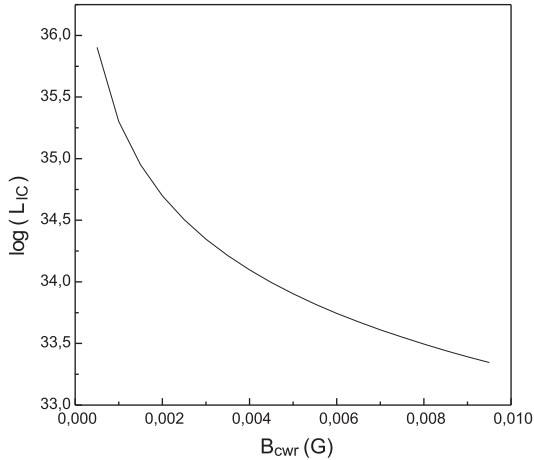}
\caption{Gamma-ray luminosity (in erg s$^{-1}$) between 100 MeV
and 20 GeV versus the magnetic field for WR 147.} \label{fig5}
\end{center}
\end{figure}

The gamma-ray luminosity due to IC scattering at the colliding
wind region, contributed by both stars, is $\sim8.0 \;10^{33}$ erg
s$^{-1}$. In Fig.~\ref{fig5} we show how this luminosity would
change with the assumed magnetic field. The gamma-ray luminosity
due to pion decay is $\sim1.0\;10^{23}$ erg s$^{-1}$ and the
corresponding to relativistic bremsstrahlung, $\sim4.7\;10^{29}$
erg s$^{-1}$.

The EGRET threshold towards the position of WR 147 is similar to
that for WR 146. A gamma-ray source at the distance of the binary
system would have been detected by EGRET if its luminosity is
greater than $2.2\;10^{34}$ erg s$^{-1}$, i.e more than a factor
two above the expected level of WR~147, according to the estimates
based on the parameters in Table~\ref{table4}. As for WR~146, this
fact could explain why no gamma-ray source was found yet towards
WR 147.

\begin{table*}
\caption[]{Gamma-ray production in the colliding wind regions of
WR 140, WR 146, and WR 147, for the energy range
100~MeV~$<$~$E$~$<$~20~GeV}
\begin{center}
\begin{tabular}{ l | l | c | c }

\hline
  & & Expected & Observed
luminosity \\Stellar system  & Mechanism & luminosity & or upper
limit$^1$
\\

           &                 &  (erg s$^{-1}$)   &    (erg s$^{-1}$)   \\

\hline &&&\\
           & Inverse Compton scattering & $\sim2.1\; 10^{34}$& \\

WR 140     & Relativistic bremsstrahlung & $\sim1.3\; 10^{31}$&
$\sim3.2\; 10^{34}$ \\

           & Neutral pion decay &   $\sim1.2\; 10^{21}$& \\
&&& \\ \hline &&&\\
           & Inverse Compton scattering & $\sim2.8\; 10^{33}$& \\

WR 146     & Relativistic bremsstrahlung & $\sim4.6\; 10^{29}$&
$<8.0\; 10^{34}$ \\

           & Neutral pion decay &   $\sim2.5\; 10^{21}$& \\
&&&\\ \hline
&&&\\
           & Inverse Compton scattering &$\sim8.0\; 10^{33}$ & \\

WR 147     & Relativistic bremsstrahlung & $\sim4.7\;
10^{29}$&$<2.2\; 10^{34}$ \\

           & Neutral pion decay &   $\sim1.0\; 10^{23}$& \\

&&&\\ \hline \multicolumn{4}{l} {$^1$ From EGRET observations and
assuming the gamma-ray sources at the same}\cr \multicolumn{4}{l}
{ distances of the stars.}\cr
\end{tabular}
\end{center}
\label{table5}
\end{table*}

\section{Discussion} \label{discussion}

\subsection{Stellar magnetic fields and the Razin--Tsytovich effect}

The magnetic field on the star surface can be computed using Eq.
(\ref{B}) and the magnetic fields estimated for the particle
acceleration regions in Sect. \ref{results}.

The rotational velocity is not well determined for the binaries
selected here. We consider typical values of 400 km s$^{-1}$ for
the WR stars, and $r_{\rm A} \sim R_*$. The resulting values for
the surface stellar fields are given in Table \ref{table6} as
$B_*$. They are between $\sim50$ and 1200 G, in accordance with
typical values estimated by Maheswaran \& Cassinelli
(\cite{mahes94}) for WR stars.

\begin{table}[]
\caption[]{Stellar magnetic fields}
\begin{center}
\begin{tabular}{ l c c }
\hline
     Star     &  $B_*$  &  $B_*^{\rm min}$ \cr
&      (G) & (G) \cr \hline &&\cr
     WR 140  &     440 & 230\cr
&&\cr

  WR 146   &     1200  & 30\cr
&&\cr

  WR 147  &        50   & 4\cr
  \hline
\end{tabular}
\end{center}
\label{table6}
\end{table}

If the magnetic field is much stronger, then WR 140 would fall
below EGRET sensitivity. On the other hand, as indicated by White
\& Chen (1995), if the field is much weaker, the Razin--Tsytovich
effect would suppress the observed synchrotron emission at cm
wavelengths. The frequency at which the radiation is suppressed
is:

\begin{equation}
\nu_{\rm R}\simeq 20 \,\,\frac{n_{\rm e}}{B}\;{\rm Hz},
\label{razin}
\end{equation}
where $n_{\rm e}$ is the electron density expressed in cm$^{-3}$.
This density is determined, in turn, by the stellar winds through:

\begin{equation}
n_{\rm e}=\frac{\dot{M}_1}{4 \pi r^2 v_{\infty,1} \mu m_{{\rm
p}}}, \label{n_e}
\end{equation}
if $m_{\rm p}$ is the proton mass.

We have taken the following minimum radio frequencies at which the
binary systems have been observed: 1.4 GHz for WR 140 (White \&
Becker \cite{whibec95}), 327 MHz for WR~146 (Taylor et al.
\cite{taylor96}), and 350 MHz for WR 147 (Setia Gunawan et al.
\cite{setia01a}). The derived minimum magnetic fields near the
shock front using Eq. (\ref{razin}) are 100 mG for WR 140, 0.6 mG
for WR 146, and 0.4 mG for WR 147. The corresponding surface
magnetic fields are given in Table \ref{table6} as $B_*^{\rm
min}$. These lower limits are quite consistent with our estimates.

During the EGRET observations towards the WR~140 system, it
occurred a periastron passage, on 1993.2. The corresponding
viewing period (212.0) lasted from March 9 to March 23, 1993, and
the measured flux was 25.7$\pm12.7\;10^{-8}$ ph cm$^{-2}$
s$^{-1}$. If we assume that all this flux is due to IC scattering
at the colliding wind region, the value of the magnetic field at
the shock region must be about 0.16 G, which leads to a stellar
surface magnetic field of $\sim$ 350 G.

\subsection{Variability}

The size of the non-thermal regions in the WR binaries considered
in this research is in the range $10^{14}-10^{15}$ cm, which is
about 3 orders of magnitude larger than the size of the stars.
Hence occultation events cannot produce significant flux changes.
Variability, instead, can result because of the changing UV photon
flux originated in very eccentric orbits and also from changes in
the wind outflow. The timescales associated with the first
process, however, exceeds the total EGRET observing lifetime by a
factor of 2, at least in WR 140, till now the only detectable
case. Future GLAST studies might reveal a modulation of the
gamma-ray emission with the orbital period for this system.
Changes in the winds could affect the injection and acceleration
rates at the shocks on shorter timescales, but these effects are
below the sensitivity of EGRET in the case of WR 140. Long term
monitoring with high-altitude ground-based GeV Cherenkov arrays
like 5@5 (Aharonian et al. \cite{ahas01}) could be very useful to
establish the changing properties of the winds.

\subsection{Additional gamma-ray contributions in the region of WR 140}

The interstellar matter surrounding the WR 140 system has been
studied by means of HI-21cm radio and IR observations (Arnal
\cite{arnal01} and references therein). Arnal found a minimum in
the neutral hydrogen distribution, built by the action of the
stellar winds, and estimated an HI mass of the surrounding
emission in $\sim 1300$ solar masses. The HI void has a major axis
of $\sim11.5$ pc and a minor axis of $\sim8.4$ pc. If particle
re-acceleration is occurring at the terminal shock of the wind,
then the material accumulated in the shell of the HI bubble could
be exposed to relativistic proton bombarding yielding an
additional contribution to the the total gamma-ray flux measured
from 3EG J2022+4317. This is a particularly interesting
possibility, since the source is classified as ``possibly extended
source" in the 3EG catalog.

The $\gamma$-ray flux expected from $p-p$ interactions is
(Aharonian \& Atoyan \cite{ahas96}):

\begin{equation}
F(E>100\;{\rm MeV})=\frac{1}{4\pi} q_{\gamma} M_{\rm HI}\; d^{-2}
m_{{\rm p}}^{-1}, \label{fp-p}
\end{equation}
where $q_{\gamma}$ is the gamma-ray emissivity of the medium whose
total mass is $M_{\rm HI}$, $d$ is the distance to the source, and
$m_{\rm p}$, as before, is the proton mass. The gamma-ray
emissivity can be related to the value observed in the vicinity of
Earth by $q_{\gamma}=k q_{\sun}$. If the proton spectrum near the
Earth is similar to the spectrum in the surroundings of WR 140 we
can approximate $k\approx k_{\rm CR}$, i.e. the enhancement factor
is given by the ratio of cosmic ray densities in the shell and in
the solar neighborhood. The latter density is usually taken as
$\sim1$ eV cm$^{-3}$, and then $q_{\sun} (E\geq100\;{\rm
MeV})\approx2.2\times10^{-25}$ $({\rm H-atoms})^{-1}$ (e.g. Dermer
\cite{dermer86}). If we consider as reasonable a cosmic ray
enhancement in the range $k\sim10 - 100$ due to re-acceleration at
the wind terminal shock and we adopt the estimate $M_{\rm
HI}\sim1300$ M$_{\sun}$ for the available mass (Arnal
\cite{arnal01}), we obtain an additional contribution to the total
gamma-ray flux from 3EG J2022+4317 of $\sim1.7 - 17 \; 10^{-8}$ ph
cm$^{-2}$ s$^{-1}$. This extended emission could be responsible
for the confusion in the EGRET detection. This flux will
contribute with a luminosity of $\sim5.7\; 10^{32}$ -- $10^{33}$
erg s$^{-1}$.

\subsection{Remarks on the electron spectrum of WR 147}

As we have mentioned in Section \ref{candi-wr147} it is not clear
that the electron injection spectrum for this system can be
represented by a canonical value $p=2$ as we have assumed for our
calculations. Even taken into account the effect of the Compton
losses it is not possible to explain the significant steepening
observed in the synchrotron flux distribution above 10 GHz
(Skinner et al. \cite{skinner99}). The radio spectrum is similar
to what would be expected from a monoenergetic electron
population. In this case, the spectral distribution of the
synchrotron radiation is (e.g. Longair {\cite{longair97}):

\begin{equation}
S(\nu)\propto \left(\frac{\nu}{\nu_{\rm c}}\right)
\int^{\infty}_{\nu/\nu_{\rm c}} K_{5/3}(\eta) d\eta,
\end{equation}

\noindent where $K_{5/3}(\eta)$ is the modified Bessel function
and $\nu_{\rm c}$ is the critical frequency near which the
emission has its maximum. In the case that the monoenergetic
electrons interact with an external photon field, the spectral
distribution of the emerging IC radiation will mimic the seed
spectrum. With the photon fields considered in this paper, the
expected gamma-ray spectrum will be similar to a black body
spectrum. In this way, gamma-ray observations of WR 147 with good
sensitivity and high spectral resolution can be used to probe the
nature of the injected relativistic particle population and to
test Skinner et al.'s (\cite{skinner99}) proposal of a
monoenergetic electron spectrum. Hopefully, GLAST will be able
solve this problem.

\subsection{Predictions for upcoming satellites}

The gamma-ray fluxes expected at other energy ranges than EGRET's,
due to IC scattering of UV photons from the secondary, are given
in Table 6. It can be seen that some INTEGRAL and GLAST detections
can be expected. The continuum sensitivity of INTEGRAL's IBIS
instrument is $2\; 10^{-7}$ ph cm$^{-2}$ s$^{-1}$ at 1 MeV for an
exposure of $10^{6}$ s. GLAST sensitivity at $E>100$ MeV for one
year survey is $\sim 4\;10^{-9}$ ph cm$^{-2}$ s$^{-1}$.
High-quality data from these instruments will help to fix the
spectral shape of the sources in different energy ranges. In
particular, notice that IBIS might observe the spectral break
predicted at $\sim2.4$ MeV for WR~140, whereas GLAST should detect
the steepening in the spectrum of WR~147 at energies $\geq$ 70
GeV.

\begin{table}[]
\caption[]{Gamma-ray flux expected at other energy ranges}
\begin{center}
\begin{tabular}{ l | c | c }
 \hline 
System & $F_{\gamma}$[15 keV --10 MeV]$^{1}$   &  $F_{\gamma}$[20
MeV -- 300 GeV]$^{2}$
\\ & (ph cm$^{-2}$ s$^{-1}$) & (ph cm$^{-2}$ s$^{-1}$)\\ \hline
&&\\
        WR 140 &  8.3$\; 10^{-4}$&    1.1$\times 10^{-6}$\cr
        WR 146 &  1.2$\; 10^{-4}$&    8.9$\times 10^{-8}$\cr
        WR 147 &  1.2$\; 10^{-3}$&    9.4$\times 10^{-7}$ \cr
&&\\ \hline \multicolumn{3}{l} {1: INTEGRAL energy range; 2: GLAST
energy range}\cr
\end{tabular}
\end{center}
\label{table7}
\end{table}

\section{Conclusions} \label{concls}

In this paper we have considered the production of gamma-ray
emission in three WR+OB binaries that are well-known non-thermal
radio sources. These systems display clear evidence of a colliding
wind zone where strong shocks are formed. The existence of a
significant synchrotron radio emission from these regions ensures
the presence of locally accelerated relativistic electrons, and
since the regions are also exposed to strong stellar photon
fields, the necessary conditions for inverse Compton production of
high-energy photons are fulfilled. We have studied here whereas,
according to the available multifrequency information, this
high-energy emission is detectable with current technology. We
conclude that in the case of WR 140, the expected flux is strong
enough as to account for an already observed but yet unidentified
EGRET source: 3EG J2022+4317. In the case of the WR stars WR 146
and WR 147 the fluxes are below the current detection thresholds,
but forthcoming experiments could detect them.

If the existence of gamma-ray emission from early-type stars can
be established beyond all observational doubt, the implications
for galactic cosmic ray astrophysics could be very important. An
important pending issue is what are the maximum energies at which
protons could be accelerated by these systems. If re-acceleration
takes place efficiently at the terminal shock fronts of the winds,
then energies above 100 TeV could be expected. The recent
detection of Cyg OB2 association by HEGRA instrument (Aharonian et
al. 2002) could be a first step towards the identification of some
stellar systems as cosmic ray sources below the so-called knee in
the galactic cosmic ray spectrum.

\begin{acknowledgements}
An anonymous referee made insightful and constructive comments on
this work. We are deeply indebted to him/her. This work has been
partially supported by CONICET (PIP 0430/98), ANPCT (PICT 98 No.
03-04881), and Fundaci\'{o}n Antorchas (PB y GER). We are grateful
to Y. M. Butt, G. Rauw, L. F. Rodr\'{\i}guez, and I. R. Stevens
for discussion.
\end{acknowledgements}


{}

\end{document}